# Near-field probing of image phonon-polaritons in hexagonal boron nitride on gold crystals

Sergey G. Menabde,[1] Sergejs Boroviks,[2] Jongtae Ahn,[3] Jacob T. Heiden,[1] Kenji Watanabe,[4] Takashi Taniguchi,[5] Tony Low,[6] Do Kyung Hwang,[3,7] N. Asger Mortensen,[2,8] Min Seok Jang[1]*

[1] School of Electrical Engineering, Korea Advanced Institute of Science and Technology, Daejeon, Korea.
[2] Center for Nano Optics, University of Southern Denmark, Odense, Denmark.
[3] Center for Opto-Electronic Materials and Devices, Korea Institute of Science and Technology, Seoul, Korea.
[4] Research Center for Functional Materials, National Institute for Materials Science, Tsukuba, Ibaraki, Japan.
[5] International Center for Materials Nanoarchitectonics, National Institute for Materials Science, Ibaraki, Japan.
[6] Department of Electrical and Computer Engineering, University of Minnesota, Minneapolis, USA.
[7] Division of Nano & Information Technology, University of Science and Technology, Daejeon, Korea.
[8] Danish Institute for Advanced Study, University of Southern Denmark, Odense, Denmark.

*Corresponding author. Email: jang.minseok@kaist.ac.kr

## Abstract

Near-field mapping has been widely used to study hyperbolic phonon-polaritons in van der Waals crystals. However, an accurate measurement of the polaritonic loss remains challenging because of the inherent complexity of the near-field signal and the substrate-mediated loss. Here we demonstrate that large-area monocrystalline gold flakes, an atomically-flat low-loss substrate for image polaritons, provide a platform for precise near-field measurement of the complex propagation constant of polaritons in van der Waals crystals. As a topical example, we measure propagation loss of the image phonon-polaritons in hexagonal boron nitride, revealing that their normalized propagation length exhibits a parabolic spectral dependency. Further, we show that image phonon-polaritons exhibit up to a twice lower normalized propagation loss, while being 2.4 times more compressed compared to the case of dielectric substrate. We conclude that the monocrystalline gold flakes provide a unique nanophotonic platform for probing and exploitation of the image modes in low-dimensional materials.

## Introduction

Direct measurement of the polariton dispersion in low-dimensional van der Waals materials is possible via near-field mapping by the scattering-type scanning near-field optical microscope (s-SNOM)(*1*). In the s-SNOM experiments, the polaritons excitation and detection are typically performed by the same nano-tip, and the recorded near-field interference pattern is due to the mode reflection at the material edge (*2-11*). Furthermore, in the case of the hyperbolic phonon-polaritons (HPP) in hexagonal boron nitride (hBN), the near-field signal carries the contribution from the HPP waves launched by the s-SNOM excitation beam at the edge of an hBN slab (*3-6, 12*).

Near-field probing of HPP can be simplified if polaritons are launched by metallic nano-particles or metal edges with a larger scattering cross-section (*4, 7, 13, 14*). In this case, the near-field interference pattern is due to the superposition of the polariton field and the quasi-uniform excitation field of the s-SNOM (*4*). However, the small size and arbitrary shape of the gold nanoparticles still lead to a diverging wavefront of arbitrary shape (*4, 14, 15*). Combination of the non-planar wavefront and the mixed near-field signal of different origin significantly complicate the near-field analysis, which often requires the development of prohibitively complex analytical models (*12, 14*).

A series of recent experimental works spotlighted a new species of low-dimensional polaritons supported by the van der Waals crystals placed in proximity to a highly conductive metal – the image polaritons, resulting from the coupling of the collective charge oscillation in the polaritonic material with their images in the metal (*16-22*). In this work, we employ large-area gold crystals to rigorously study the dispersion of hyperbolic image phonon-polaritons (HIP) in hBN slabs. The well-defined, ~20 µm-long crystalline gold edges efficiently launch HIP with a planar wavefront (Fig. 1A), significantly simplifying the near-field analysis. Furthermore, monocrystalline gold has an atomically flat surface with RMS roughness (*23, 24*) as small as 1 Å (Fig. S1), which, along with the crystalline nature of hBN, eliminates roughness-mediated scattering of the propagating polaritons.

Taking advantage of the physical properties of the gold crystals, we accurately measure the complex propagation constant of the HIP. Especially, the propagation loss (or the imaginary part of the propagation constant) of HIP is quantitatively determined in this study for the first time. Our experimental data perfectly agrees with the analytically calculated loss in a system without scattering. Furthermore, due to the scattering-free propagation, we detect a high-order image mode with momentum of $5.9 \times 10^5$ cm$^{-1}$ even in a relatively lossy (naturally abundant) hBN crystal. So far, phonon-polaritons with such high momentum have been observed only in the low-loss isotopically enriched hBN (*6*). Finally, we show that the fundamental image mode exhibits both stronger field confinement and longer normalized propagation length (in optical cycles) compared to the fundamental mode in the hBN on dielectric substrate, notably similar to the behavior of image graphene plasmons (*16*).

## Results

hBN is an anisotropic van der Waals crystal which supports propagating HPP in the two reststrahlen bands where the in-plane ($\varepsilon_{xy}$) and out-of-plane ($\varepsilon_z$) components of the permittivity tensor have opposite sign of the real part (*3, 25*). We restrict our study to the second reststrahlen band (1370–1610 cm$^{-1}$; Re{$\varepsilon_{xy}$} < 0, Re{$\varepsilon_z$} > 0) which is accessible by the s-SNOM coupled with a quantum cascade laser.

Due to the hyperbolic dispersion, propagation of phonon-polaritons in hBN is restricted to a certain direction defined by the ratio $\varepsilon_{xy}/\varepsilon_z$, which leads to the quantization of the waveguide modes according to the linear scaling rule $k_z t \sim l$, where $k_z$ is the out-of-plane component of the polariton wavevector, *t* is the hBN slab thickness, and *l* = 1,2,3… is the mode order (*25, 26*). Figure 1B shows the field profile of the first- and the second-order HPP modes (HPP$^1$ and HPP$^2$, respectively), numerically calculated by the full-wave simulations. Due to the "reflection" by image charges, the first-order HIP mode (HIP$^1$) is equivalent to the symmetric HPP$^2$ in a twice thicker hBN slab. Despite the tighter field confinement, the electric field of the HIP mode penetrates into the air above hBN at a distance ~ $t$ and thus can couple to the s-SNOM nano-tip acting as a $z$-oriented electric dipole (Fig. 1C).

We obtain near-field images of an hBN flake on top of the hexagonal gold crystal on a borosilicate crown (BK7) substrate (Fig. 1D). Area "1" marks the sample structure where the illustrative near-field image of Fig. 1E has been obtained, and areas "2" and "3" show where the HIP and HPP dispersions have been measured, respectively (Fig. 2).

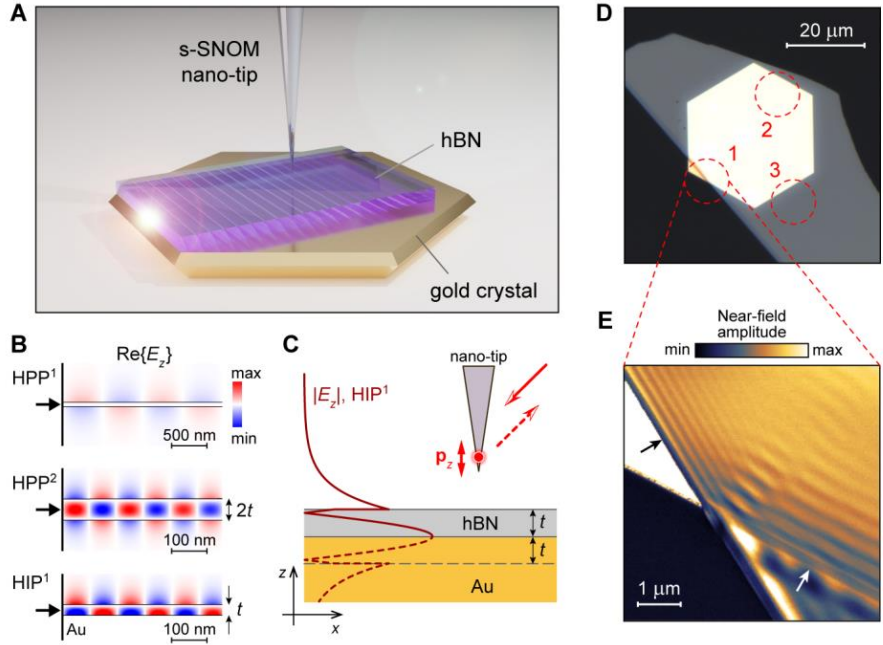

**Figure 1. Near-field probing of image phonon-polaritons in hBN.** (**A**) Atomically-smooth monocrystalline gold flake with ~10 μm-long edges provides a precise experimental platform for near-field probing of image polaritons. (**B**) Numerically calculated field profile of the first- and second-order HPP$^{1,2}$ in a suspended hBN slab of thickness $2t = 60$ nm, and the first-order HIP$^1$ in hBN of thickness $t = 30$ nm on gold; modes are launched at the port boundary condition shown by the black arrow, carrying the same power. (**C**) Field amplitude $|E_z|$ profile of the HIP$^1$ eigenmode (solid) and its mirror image in gold (dashed), equivalent to field profile of the second-order HPP mode in a twice thicker hBN slab. (**D**) Optical image of a 57 nm-thick hBN flake on a gold crystal with marked areas of near-field experiments: "1" corresponds to the illustrative near-field image shown in (**E**), "2" and "3" is where the HIP and HPP dispersion and FOM have been measured, respectively. (**E**) Near-field image of a 63 nm-thick hBN slab cross-covering the edge of a gold crystal. Black arrow indicates the hBN edge on gold where HIP interference pattern is predominantly formed by the tip-launched modes reflected from the hBN edge, corresponding to the standing wave profile. White arrow indicates the gold crystal edge under the hBN where HIP interference pattern is predominantly formed by the interference of HIP and the incident quasi-uniform excitation beam.

In the absence of other scatterers, the HIP are efficiently launched by the s-SNOM tip and form a standing wave interference with the reflected mode. At the same time, metallic particles efficiently launch HIP when placed underneath or on top of the hBN slab due to a much larger scattering cross-section (*4*). Furthermore, since the slab is continuous in the latter case, the HIP launched by the tip does not experience strong reflection from the metal particles and mostly propagate away (see Supplementary Materials), thus the near-field interference pattern corresponds to the superposition of the HIP and the quasi-uniform illumination beam (*4*). The near-field image of an hBN slab crossing the crystalline gold edge (Fig. 1E) demonstrates the two cases of HIP excitation: black arrow indicates the hBN edge on top of gold, and the interference pattern of the standing wave with period of $\lambda_{HIP}/2$, where $\lambda_{HIP}$ is the HIP wavelength. White arrow indicates the crystalline gold edge under the hBN, and the interference pattern formed by the HIP and the illumination beam with period of $\lambda_{HIP}$.

Because polaritons are launched by the straight and long ($\approx 20$ μm $\gg \lambda_{HIP}$) crystalline gold edge (areas "2" and "3" in Fig. 1D) and experience no scattering, the near-field interference pattern directly corresponds to the decaying field amplitude $|E_z|$ of the propagating modes with a planar wavefront. Therefore, Fourier transform of the near-field interference fringes readily provides value of the complex propagation constant, as explained below.

Due to the planar wavefront of observed phonon-polaritons, it is possible to integrate fringe profiles across the large imaging area, which significantly improves the data quality. Near-field interference patterns at different excitation frequency and the corresponding integrated fringe profiles with removed background are shown in Fig. 2A,C, and the corresponding Fourier spectra (black) are shown in the adjacent panels in Fig. 2B,D (see Fig. S4 for data at 1570 and 1450 cm$^{-1}$). The peak position of the fitted Lorentzian (red) gives the real part, and its FWHM provides the imaginary part of the propagation constant. The black arrows in Fourier spectra indicate the double frequency of the HIP[1] and HPP[1] peaks where the signal from the tip-launched standing wave is expected. Indeed, a very weak double-frequency signal is present in all HIP spectra, but its small amplitude indicates that fitted Lorentzian profiles correspond solely to the modes launched by the gold edge. Extracted wavenumbers for the HPP and HIP modes are shown as red circles in Fig. 3A and B, respectively.

Furthermore, a weak Fourier signal is visible near the analytically predicted frequency of the ultra-confined HIP[2] mode at 1480 and 1510 cm$^{-1}$ (blue arrows in Fig. 2B); the HPP[2] mode remains undetected in our experiments. Extracted wavenumbers of the HIP[2] mode are shown as red squares in Fig. 3B. The second-order image mode corresponds to the fourth-order HPP mode, which has been reported only once in the ultra-low-loss isotopically enriched hBN slabs (*6*). Excitation of the ultra-confined HIP[2] even in the relatively lossy (naturally abundant) hBN, as the one used in our experiments, highlights the absence of scattering loss in the sample.

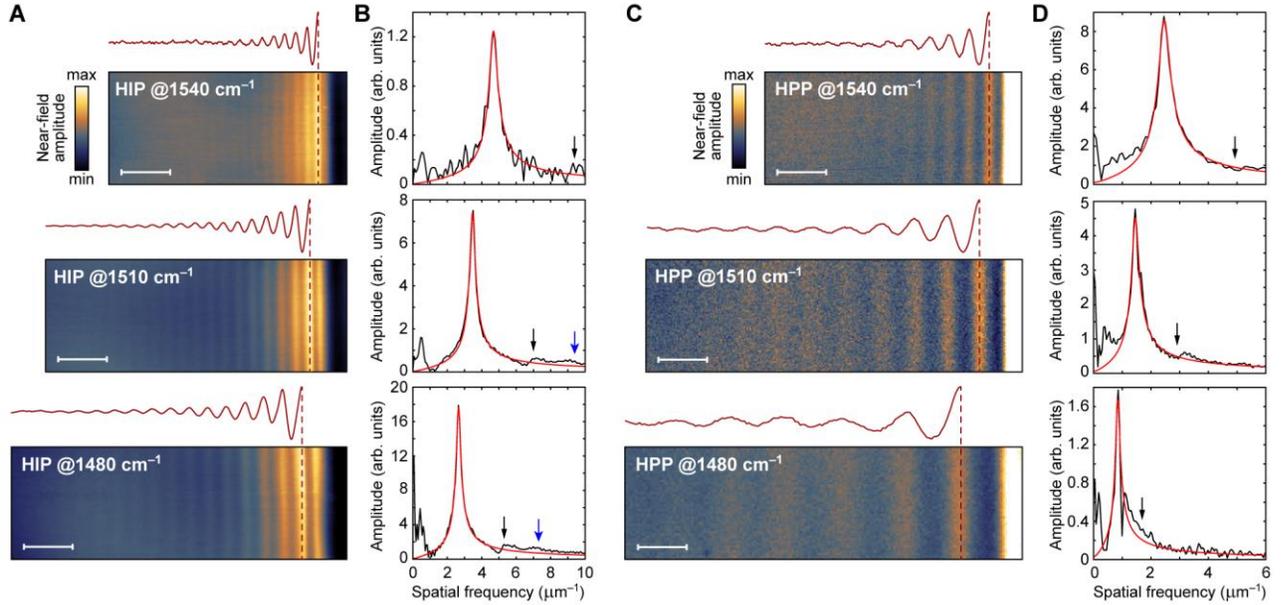

**Figure 2. Fourier analysis of the near-field interference fringes.** (**A**) Near-field images of the hBN slab on gold crystal near the crystal edge (area "2" in Fig. 1D), with clearly visible HIP fringes mapped at different excitation frequency. Corresponding integrated fringe profiles with removed background are shown above each near-field image. (**B**) Fourier spectra (black) of the integrated fringe profiles in (**A**), and the Lorentzian fit (red). Black arrow indicates the double frequency where the signal from the tip-launched HIP is expected; blue arrow indicates the analytically predicted frequency of the second-order HIP mode. (**C,D**) Same as in (**A,B**) for the HPP modes launched by the gold edge and propagating in hBN on glass substrate (area "3" in Fig. 1D). All scale bars are 1 μm.

As a figure of merit (FOM) of polariton damping, we use the normalized propagation length in optical cycles given by $\text{Re}\{k_x\}/(2\pi\text{Im}\{k_x\})$, where $k_x$ is the propagation constant. First, parameters of the hBN dielectric function are determined from the Raman spectroscopy of our samples and the previously reported data (*6*) (see Supplementary Materials). Using the recovered hBN dielectric function, we analytically calculate the dispersion (color map in Fig. 3A,B) and FOM (curves in Fig. 3C,D) of the HPP and HIP modes. Interestingly, FOM of the $HPP^2$ mode exhibits a parabolic spectral dependency, maximizing at 1510 cm$^{-1}$ where its propagation length is 1.65 times larger than that of the $HPP^1$ mode (Fig. 3C). Naturally, similar FOM is obtained for the $HIP^1$ mode (Fig. 3D). However, the FOM of the $HIP^2$ mode has only a slightly better maximal value. The drastic FOM difference between the $HPP^1$ and $HIP^1$ modes can be explained by a much smaller group velocity of the compressed HIP, while the polariton lifetime (material loss) practically does not change. The same dispersion property has been reported for the highly-compressed image graphene plasmons (*16, 18*). Besides, due to the linear scaling of the wavevector with *t*, HIP FOM practically does not depend on the hBN thickness.

Experimentally obtained values of the FOM for the fundamental HIP mode are shown by red circles in Fig. 3D, in excellent agreement with the analytical predictions based on the recovered dielectric function of hBN. We note that our method provides an accurate value of loss only for the image modes since gold can be approximated as a perfect electric conductor at mid-IR frequencies. The possible presence of a few nm-thick air gap between the gold and hBN does not affect the FOM (see Supplementary Materials). However, given the perfect agreement between the measured and calculated (without the gap) HIP momentum (Fig. 3B), we conclude that the thickness of the gap in the area of measurements is negligibly small.

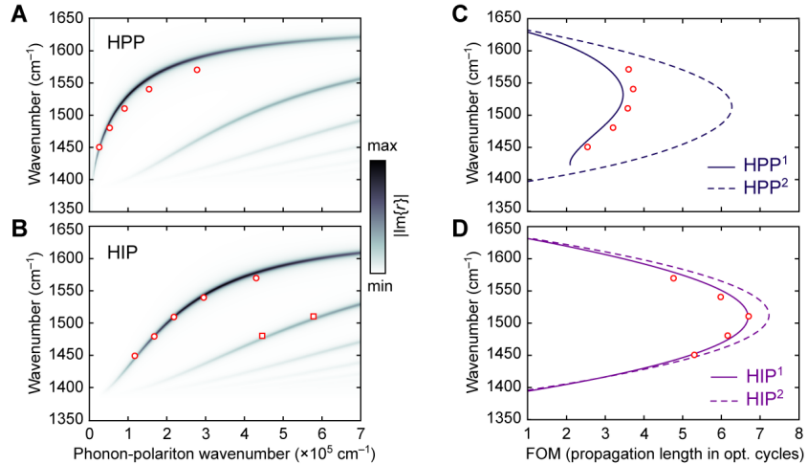

**Figure 3. Dispersion and FOM of HPP and HIP modes.** (**A**) Dispersion of the HPP modes calculated for a 57 nm-thick hBN slab on BK7 glass substrate (color map), and measured by near-field probing (red circles). (**B**) Calculated (color map) and measured (red data points) dispersion of the HIP in a 57 nm-thick hBN slab on monocrystalline gold flake. (**C**) FOM of the first- and second-order HPP modes, calculated for the case in (A) (curves), and measured by near-field probing (red circles). (**D**) Calculated (curves) and measured by near-field probing (red circles) FOM of the fundamental HIP mode which dispersion is shown in (B).

Experimentally measured maximal FOM in our sample is 3.72 for the HPP (at 1540 cm$^{-1}$), and 6.72 for the HIP (at 1510 cm$^{-1}$), revealing a 1.8 times difference. When compared at the same excitation frequency of 1510 cm$^{-1}$, the HIP FOM is $\approx$ 1.9 times larger than that of the HPP, while having 2.4 times shorter wavelength ($\lambda_{HIP}$ = 287 nm; $\lambda_{HPP}$ = 689 nm).

It is important to note that the damping of phonon-polaritons in hBN strongly depends on its isotopic composition. Giles et al. reported a greatly improved FOM of the HPP in the isotopically pure hBN, with the largest FOM $\approx$ 6.7 measured in $^{10}$B 98.7% sample (*6*). Using the provided dielectric function for such isotopically pure hBN, we analytically predict an exceptionally large FOM = 19 for the HIP$^1$ mode (calculated for 100 nm-thick hBN at 1520 cm$^{-1}$).

In order to estimate the impact of the gold roughness on HIP scattering loss, we have conducted the near-field probing of HIP on evaporated gold (RMS surface roughness $\approx$1 nm) in the hBN flakes of different thickness. In this case, only the standing wave interference fringes are available for analysis (fig. S6), therefore we also obtain similar near-field interference images in 30 and 63 nm-thick hBN on a gold crystals, shown in Figs. 4A and C. The adjacent Figs. 4B and D show the near-field maps of 22 and 66 nm-thick hBN on evaporated gold. It is immediately evident that the roughness-mediated scattering on evaporated gold compromises the near-field probing of HIP in thinner hBN sample, while its effect is less pronounced in thicker hBN where $\lambda_{HIP}$ is three times larger. At the same time, negligible roughness of the crystalline gold does not contribute to HIP damping when *t* is reduced from 63 to 30 nm (Fig. 4A,C).

We obtain the approximate FOM of the HIP on evaporated gold by analyzing the Fourier spectra of the interference fringes. In this case, an additional correction is required to compensate for the circular diverging wavefront and the double frequency of the fringes, while assuming a perfect HIP reflection at the hBN edge (see Supplementary Materials). Results of this approximate analysis are shown in Fig. 4E, correctly predicting the general tendency of the analytical FOM (grey curve), but only in the 66 nm-thick hBN (blue open circles), while HIP in the 22 nm-thick slab propagate 20% less optical cycles at maximal FOM. Stronger HIP damping in thin hBN on evaporated gold can be understood from the HIP field amplitude distribution upon its reflection from the edge (Fig. 4F), where |E| is maximal immediately next to the gold surface. This, along with the shorter $\lambda_{HIP}$, leads to a stronger HIP scattering by surface roughness.

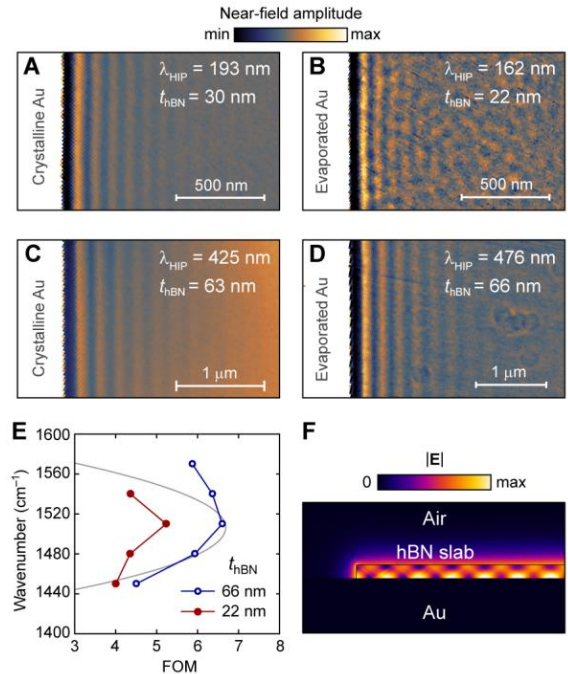

**Figure 4. Roughness-mediated HIP scattering on evaporated gold.** (**A-D**) Near-field scans of hBN slabs of different thickness (as noted) near the hBN edge: (**A,C**) on crystalline gold; (**B,D**) on evaporated gold. Excitation frequency is 1480 cm$^{-1}$ in all cases. (**E**) Normalized propagation length of HIP in optical cycles, approximated from the near-field scans of hBN on evaporated gold: for $t = 22$ nm (red filled circles) and $t = 66$ nm (blue empty circles). Grey line shows the analytical solution for the HIP FOM. (**F**) Electric field amplitude profile of the HIP mode propagating to and reflecting from the hBN edge, numerically calculated by the full-wave simulations.

## Discussion

In conclusion, we demonstrate that the monocrystalline gold flakes provide a platform for precise near-field probing of image polaritons, where modes with planar wavefront are launched by the long crystalline gold edges and propagate on an ultra-flat low-loss substrate. As a topical example, we measure the complex propagation constant of HIP in hBN within the second reststrahlen band. Our experimental data precisely follows the analytical prediction for the independently recovered hBN dielectric function. We demonstrate the larger field confinement and simultaneously longer normalized propagation length of the HIP modes compared to the HPP, notably similar to the behavior of the image graphene plasmons. This unique property potentially allows the combination of strong light-matter interaction and wave phenomena within a single nanophotonic platform.

## Materials and Methods

### Sample preparation

Monocrystalline gold flakes were synthesized using the modified Brust–Schiffrin method (*27*) via thermolysis (*28*). An aqueous solution of the chloroauric acid ($HAuCl_4 \cdot 3H_2O$ in concentration 5 mmol/L) was mixed with a solution of tetraoctylammonium bromide (TOABr) in toluene and stirred for 10 minutes at 5000 RPM. Then the mixture was left to rest for approximately 10 minutes for the separation of aqueous and organic phases. The BK7 glass substrate was prepared by pre-cleaning in ultrasonic bath in acetone, isopropyl alcohol (IPA), and ultrapure water (Milli-Q). After blow-drying by nitrogen gas, the substrate was baked on a hot plate at 200 °C for approximately 5 minutes for dehydration. Then, a few microliters of the organic phase were drop-casted onto a substrate which was then left on the hot-plate at 130 °C for 24 hours. After that, the sample was cleaned in toluene at 75 °C, acetone, and IPA, which removed most of the organic solvent.

hBN nano-flakes were mechanically exfoliated from the bulk single crystal and transferred by the polydimethylsiloxane stamp on top of the gold crystals.

### Sample characterization

Near-field images were obtained by commercial s-SNOM (Neaspec GmbH) coupled with the tunable quantum cascade laser (Daylight Solutions, MIRcat). The Pt-coated AFM tips (Nano World, ARROW-NCPt) were used with tapping frequency $\Omega$ around 260 kHz and oscillation amplitude 60-70 nm in a non-contact mode. The background-free interferometric signal (*29*) demodulated at third harmonic $3\Omega$ was used to generate all near-field images. Samples were oriented in such a way that the plane of incidence of the s-SNOM illumination beam was never orthogonal to the gold edge, which significantly decreased the near-field background due to the strong scattering at the atomically-sharp edge. Thickness of the hBN flakes was measured in the same AFM tapping mode as used for near-field imaging.

Raman spectroscopy of hBN on gold was performed using the LabRAM HR Evolution Visible-NIR system (HORIBA). Permittivity of BK7 glass substrate at mid-IR frequencies was measured by ellipsometry (J.A. Woollam, IR-VASE).